\documentclass[journal]{IEEEtran}
\usepackage{amsmath,amsthm,amsfonts,amssymb,latexsym}
\usepackage{times}
\usepackage{helvet}
\usepackage{courier}
\usepackage{fancyhdr,graphicx}
\usepackage{tabularx}
\usepackage{subcaption}
\usepackage{xcolor}
\usepackage{multirow}
\usepackage{xcolor,colortbl}
\newtheorem{Theorem}{Theorem}

\newtheorem{Definition}[Theorem]{Definition}
\begin{document}
\title{Fair Pricing In Heterogeneous Internet of Things Wireless Access Networks Using Crowdsourcing}
\author{\IEEEauthorblockN{Vahid Haghighatdoost, Siavash Khorsandi and
 Hamed Ahmadi
}
\thanks{Vahid Haghighatdoost and Siavash Khorsandi are with Computer Engineering and IT Department, Amirkabir University of Technology, Tehran, Iran \{haghighatdoost, khorsandi\}@aut.ac.ir. Hamed Ahmadi is with the Department of Electronic Engineering, University of York, UK, hamed.ahmadi@york.ac.uk.
}
}
\IEEEtitleabstractindextext{%
\begin{abstract}
Price and the quality of service are two key factors taken into account by wireless network users when they choose their network provider. The recent advances in wireless technology and massive infrastructure deployments has led to better coverage, and currently at each given wirelessly covered area there are a few network providers and each have different pricing strategies. These providers can potentially set unfair expensive prices for their services. In this paper, we propose a novel crowdsourcing-based approach for fair wireless service pricing in Internet of Things (IoT).
In  our  considered  oligopoly,  the  regulatory  sets a dynamic maximum allowed price of service to prevent anti-trust behaviour and unfair service pricing. We propose a three-tire pricing model where the regulator, wireless network providers and clients are the players of our game. Our method takes client preferences into account in pricing and discovers the fair service pricing just above the marginal costs of each network provider. Our results show that our model is not prone to collusion and will converges only if one network announces the fair price. 

\end{abstract}

}

\maketitle
\IEEEdisplaynontitleabstractindextext
\IEEEpeerreviewmaketitle

\section{Introduction}
In a typical urban environment a mobile user can choose to connect to different network providers using different technologies. This includes various generations of mobile and cellular networks such as 3G, LTE and 5G technologies, various versions of WiFi networks, and cognitive radio networks (CRNs) which has led to the concept of heterogeneous wireless access network (HWAN) where all these technologies and networks are available for clients and spectrum assignment needs to be efficiently managed \cite{Ref_1}. Regulatory bodies, in this environment, usually play a pivotal role in leading the system towards optimal operating point, both in terms of spectrum allocation to competing networks and also in terms of pricing \cite{Ref_3}. This environment becomes even more complicated with presence of Internet of Things (IoT) devices/users.

Although there has been extensive works in dynamic spectrum access (DSA) techniques \cite{Ref_6,Ref_7,Fungible},
spectrum licenses are still usually issued for a predefined period of time and currently are not reassigned in short time scales. Therefore, in a typical condition where the number of available Wireless Network Providers (WNPs) is limited and spectrum allocations cannot be dynamically modified, price regulation will play an important role to both guarantee a healthy profit margin for the WNPs and prevent unfair pricing and anti-trust behaviors such as price-fixing. 

In an oligopoly market where the number of WNPs and the available spectrum are limited, Walrasian schemes do not apply as the rules of supply and demand equilibrium leads to grossly unfair pricing because it is not a full competitive market. A full competitive  market is a market structure where a large number of buyers and sellers are present, and all are engaged in buying and selling of semi-homogeneous products at prices prevailing in the market \cite{Ref_8}. Basically, in an oligopoly market, unlike a full competitive market, user response to providers' unfair pricing does not necessarily lead to low revenue demotivating this behavior. Therefore, when WNPs are able to announce their price, the regulator needs to intervene to prevent anti-trust behavior and unfair service pricing. Commonly used auction-based pricing strategies are not effective in such scenarios \cite{Ref_6,Ref_7,Ref_9}, since these schemes are designed to maximize the auctioneer’s utility and not the clients’ welfare. Best response techniques need the knowledge of providers’ marginal cost which is not readily available \cite{Ref_4}. We also need to take client preferences into account in pricing since WNPs’ weights from clients’ point of view depend on both cost and quality of service (QoS). As the service quality of various WNPs may vary, their service prices may also be different accordingly. This makes the price regulation decisions very challenging as quality-cost trade off from clients’ point of view are not known to the regulatory. To the best of our knowledge, there is no existing study that addresses these issues. This is based on our systematic search on relevant phrases like ``Wireless Access Fair Pricing" in popular scientific articles search engines, repositories and major publishers like IEEE, ACM and Springer.

This work fills the existing gap in the field by providing a framework and attributed mechanisms for implicit discovery of marginal costs through crowdsourcing. It does not suffer from the constraints and limitations of the existing schemes and is robust in reacting to selfish behaviors. Our proposed mechanism discovers the fair service pricings, just above the marginal costs of the WNPs.  This is performed through interaction between the regulator and the clients. 

In a number of previous works, the regulator is considered as a stakeholder aiming to maximize its direct income from licensing fees. However, as pointed out in \cite{Ref_1} and in \cite{Ref_2}, the radio spectrum is a major national asset, contributing a significant value  to  the  economy  and  playing  a  vital  role  in  social welfare  and  national  security. As emphasized in \cite{Ref_1} the spectrum allocation should be Pareto efficient meaning that the regulator is responsible for spectrum redistribution in such a way that those who are made better off by the redistribution could fully compensate those who are made worse off prior to the redistribution. We, therefore, consider the regulator as a neutral arbiter whose main mandate is to maximize the social welfare rather than its direct income.
Our main contributions in this work are as follows:
%\textcolor{blue}{
\begin{itemize}
	\item We develop a realistic model for preparing request bundles for clients based on microeconomics. In our model the clients generate their request bundles by maximizing their utility which is a function of WNPs' weights subject to their budget constraint and price of each WNP. WNP's weights are internal information of each client which are dynamically updated based on the price of the WNP's service and its quality of service.
	\item We crowdsource the clients beliefs on WNPs' fairness and use it in revising WNPs' price caps. The regulator in our mechanism aims to maximise the social welfare by maintaining a price cap for WNPs. Depending on clients' crowdsourced feedback the regulator increases or decreases this price cap. This feedback mechanism does not exist in previous works on dynamic games. This feedback mechanism provides extra information about WNPs pricing strategy and quality of provided service which enables the regulator to make proactive decision in regulating the market. This additional information changes the mechanism in a way that has not been studied in the literature. It also enables the mechanism to be robust in detecting behaviour changes and react to it. 
	\item We present an adaptive punishment and reward for WNPs according to the crowdsourced feedback from the clients. The price cap for each WNP is reviewed dynamically based of the crowdsourced data. However, the rate of changes in the cap depends on the level of honesty and/or unfairness of the WNPs. 
	\item We show via simulation that this system is capable of reacting to dynamic behavior of WNPs. In our simulations we experiment different scenarios and test the system under probabilistic honest/unfair pricing of WNPs over time. Our results show that the system converges to the fair prices only if there is one WNP which is honest most of the time (high probability of honesty). We also analyse the complexity of the proposed mechanism. 
\end{itemize}
%}

The rest of this paper is organized as follows. Section \ref{Sec:Rel} contains a review of the related works. Our system model and problem formulation is presented in Section III. We explain the proposed Crowdsourcing Price Control (CSPC) scheme in section IV. We analyze the convergence of the system and presented the numerical results Section V. Finally, Section VI contains our concluding remarks.

\section{Related Works}\label{Sec:Rel}

\subsection{Pricing Objectives}
Service pricing is addressed in various contexts in wireless/mobile access networks. The WNPs that are licensed to use a predefined spectrum have considered many ways regarding pricing criteria to achieve a higher amount of income.  In most cases, these schemes are simple flat-rate pricing  \cite{Ref_10}. In \cite{Ref_9}, a dynamic pricing method is developed that considers Service Level Agreement (SLA) for clients in pricing and proposes different prices for different categories of SLAs. There are also smart pricing methods such as \cite{Ref_11,Ref_12} in which the WNP sets the price based on connection duration, or other WNP's parameters. In \cite{Ref_13} the volume-based service pricing for cellular networks is presented where they tried to categorize different service bundles and announce a price for each category according to data volume size, user budget, data rate, and service blocking probability. In most pricing approaches, the goal is to maximize WNPs' income \cite{Ref_6,Ref_7}. In works like \cite{Ref_14}, pricing is used to increase the usage of available spectrum rather than WNP’s satisfaction.
Pricing is also addressed in various works in the field of Cognitive Radio Networks (CRNs) where licensed primary networks (PUs)  act as WNPs and secondary networks (SUs) act as clients that can lease the spectrum bands from PUs \cite{Ref_15}. The idea in these works is to both maximize spectrum utilization and PUs’ revenues. None of them, however, addressed the issue of fair pricing. In the area of IoT pricing mechanisms are studied mainly in the contexts of sensed data pricing rather than wireless data pricing \cite{luong2016data}.

\subsection{Pricing Strategies}
One of the most popular solutions for pricing and resource allocation is the auction mechanism where the regulatory is considered to be the auctioneer with the goal of selling the commodity (spectrum) at the highest price, not the fairness of final service price that client pays (\cite{Ref_6,Ref_16,AhmadiMag}). Authors in \cite{Ref_6}, formulate a balance between utility of the auctioneer and WNPs as bidders aiming to maximize the combined utility of bidders and the auctioneer. To reduce the interference and reuse channels, Zhu et al. in \cite{Ref_16} propose a simple heuristic auction for spectrum allocation in multi hop WNPs to clients. Their auction mechanism guarantees both truthfulness and interference-free channel allocation. %Their goal is reducing the interference and reuse of channels.

Another group of works have followed game-theoretic approaches and considered the regulatory as a selfish player. Chen et al. in \cite{Ref_17} proposed a three tier game model to balance the income of the Federal Communications Commission (FCC) and the aggregate utility of clients. They considered the income of FCC as important as the utility of clients and WNPs. In \cite{Ref_15}, authors proposed a two-stage Stackelberg game model for CRNs, where in first stage, PUs announce their price and in second stage, SUs prepare their request bundle. Three-stage Stackelberg games have been studied for resource management in IoT fog network \cite{Stack3} and mobile data market \cite{Stack2}.  Although Stackelberg-based approaches can nicely model the behavior of the leader and follower(s), they are inefficient in our work. In our work a WNP, leader, can increase its price of service unfairly as the number of WNPs are limited and clients are forced to pay unfair prices since all of clients' requests could not be covered by other competing WNPs. 

Crowdsourcing has been considered recently in wireless networks for cooperative system design. In \cite{Ref_18}, authors proposed a general incentive mechanism to improve the efficiency and utility of the mobile crowdsourcing system. The information gathered from the clients is considered to be any information of value for the WNPs. In \cite{Ref_18}, the use of crowdsourcing techniques for managing QoE in mobile networks is discussed. 

\begin{table}[]
\caption{Literature classification summary}\label{Tbl:A}
\begin{tabular}{|p{0.20\columnwidth}|p{0.45\columnwidth}|p{0.20\columnwidth}|}
\hline

\multirow{3}{*}{Pricing model}     & Flat rate                                        & \cite{Ref_10} \\ \cline{2-3} 
                                   & Dynamic rate                                     &  \cite{Ref_9}\\ \cline{2-3} 
                                   & Smart pricing                                    & \cite{Ref_11, Ref_12, Ref_13} \\ \hline
\multirow{3}{*}{Pricing objective} & Maximize WNP income                              & \cite{Ref_6,Ref_7} \\ \cline{2-3} 
                                   & Maximize Spectrum utilization                    & \cite{Ref_14} \\ \cline{2-3} 
                                   & Jointly maximise income and spectrum utilization &  \cite{Ref_15}\\ \hline
\multirow{3}{*}{Regulator role}    & Regulator as an auctioneer                          & \cite{Ref_6,Ref_16} \\ \cline{2-3} 
                                   & Regulator as a selfish player                      & \cite{Ref_17,Ref_15,Stack2} \\ \cline{2-3} 
                                   & Regulator maximizing the social welfare          & \cite{Ref_18} \\ \hline
                                   
\end{tabular}
\end{table}

In \cite{Ref_4}, we assumed that WNPs announce fair prices and we proposed a mechanism for the regulator to adjust the spectrum allocation to maximize the social welfare. We considered the clients as well as WNPs satisfaction in our social welfare formulation. Consideration of fair pricing for WNPs was very optimistic, but in current work we focus on a mechanism that guides the WNP to fair pricing. In \cite{Ref_5}, we considered that, there are only two WNPs (3G and 4G Network providers) that are enforced to sell the prepared service, otherwise, they will be penalized by a regulator and the regulator reduced their spectrum amount. Hence WNPs were trying to attract clients by reducing their price of service to the extent that they do not harm. We used best response solution to solve the network-to-network and network-to-client competitions and this solution when number of WNPs exceed from two is very difficult and adjusting the spectrum allocation table is not applicable because the lincenses are usually issued annually. In this paper, we use a crowdsourcing system to evaluate WNPs and determine whether the announced price by a WNP is fair or not. Table \ref{Tbl:A} summarises the classification of studied literature based on their pricing strategy and the role of the regulator.

\begin{table}
\caption{Table of variables and acronyms}\label{Tbl:1}
\begin{tabular}{|p{0.15\columnwidth}|p{0.75\columnwidth}|}
\hline
{\centering \textbf {Parameter}} & {\centering \textbf{Description}}\\
\hline
\centering{$N$} &{Number of WNPs in HWAN}\\
\hline
\centering{$M$} &{Number of clients in HWAN}\\
\hline
\centering{$r_j^i$} &{Requested bit-rate of client  $i$ from WNP  $j$ at current iteration}\\
\hline
\centering{$\boldsymbol{r^i} (t)$} &{The requested bit-rate vector of client $i$ from $N$ existing WNPs at iteration $t$ at each PCC.$\boldsymbol{r^i} (t)=(r_1^i,\dots,r_N^i)_t$}\\
\hline
\centering{$x_j^i (t)$} &{Assigned bit-rate to client $i$ by WNP $j$ at iteration $t$ at each PCC}\\
\hline
\centering{$\boldsymbol{x^i}$} &{Assigned bit-rate to client $i$ from WNP $j$, $\boldsymbol{x^i}=(x_1^i,\dots,x_N^i)$}\\
\hline
\centering{$x_j^i$} &{Assigned bit-rate to client $i$ from WNP $j$}\\
\hline
\centering{$\boldsymbol{x^i}(t)$} &{The assigned bit-rate vector to client $i$ by $N$ existing WNPs at iteration $t$ at each PCC. $\boldsymbol{x^i} (t)=(x_1^i,\dots,x_N^i)_t$}\\
\hline
\centering{$\boldsymbol{x^{(i,f)}}$} &{Final assigned bit-rate vector to client $i$ by $N$ existing WNP at PCC $f$.$ \boldsymbol{x^{(i,f)}}=(x_1^{(i,f)},\dots,x_N^{(i,f)})$}\\
\hline
\centering{$\boldsymbol{p}$} &{The vector of announced prices of all WNPs at current PCC. $\boldsymbol{p}=(p_1,\dots,p_N)$}\\
\hline
\centering{$\boldsymbol{\tilde{p}}$} &{The ceiling of prices or maximum allowed price for a unit of the service of WNPs defined by the regulatory at current PCC. $\boldsymbol{\tilde{p}}=(\tilde{p}_1,\dots,\tilde{p}_N )$}\\
\hline
\centering{$\boldsymbol{\tilde{p}}(f)$} &{The ceiling of prices or maximum allowed price for a unit of the service of WNPs defined by the regulatory at price controlling cycle $f$. $\boldsymbol{\tilde{p}}(f)=(\tilde{p}_1(f),\dots,\tilde{p}_N(f))$}\\
\hline
\centering{$\boldsymbol{\rho}^i$} &{The suitable ratio of prices (SRP) from point of view of client $i$. $\boldsymbol{\rho}^i=(\rho_1^i,\dots,\rho_N^i)$}\\
\hline
\centering{$\bar{P}$} &{The average of all announced prices of all WNPs in the market. $\bar{P}=\frac{1}{N} \sum_{j=1}^N p_j$}\\
\hline
\centering{$\boldsymbol{\hat{p}}^i$} &{estimated fair price of all WNPs by client $i$ or the suitable price from the sight of client $i$. $\boldsymbol{\hat{p}}^i=(\hat{p}_1^i,\dots,\hat{p}_N^i )$
$(\boldsymbol{\hat{p}}^i=\bar{P}.\boldsymbol{\rho}^i)$ } \\
\hline
\centering{$TC_j$} &{Total cost function, determines the total cost of WNP $j$ for preparing the load $L_j$. $TC_j=TC_j (L_j )$} \\
\hline
\centering{$MC_j$} &{Marginal cost function, determines the marginal cost of WNP $j$ for preparing the load $L_j$. $MC_j=MC_j (L_j )$} \\
\hline
\centering{$\bar{MC}$} &{The average of all marginal costs $\frac{1}{N} \sum_{j=1}^N MC_j$ for current prepared service by WNPs} \\
\hline
\centering{$\boldsymbol{s^i}$} &{The prepared crowdsourcing data by client i to send for the regulatory.$\boldsymbol{s^i}=(s_1^i,\dots,s_N^i)$} \\
\hline
\centering{$S_j $} &{Total perfect request for the prepared service by WNP $j$ that calculated from crowdsourcing data $(S_j=\sum_{i=1}^M s_j^i )$} \\
\hline
\centering{$\boldsymbol{\tilde{w}^i}$} &{The initial weight of $N$ WNPs as the sight of client $i$. $\boldsymbol{\tilde{w}^i}=(\tilde{w}_1^i,\dots,\tilde{w}_N^i )$} \\
\hline
\centering{$\boldsymbol{w^i}$} &{The adjusted weight of WNPs as the sight of client $i$ at current price controlling cycle.$\boldsymbol{w^i}=(w_1^i,\dots,w_N^i) $} \\
\hline
\centering{$H^i$} &{The budget of client $i$ } \\
\hline
\centering{$R^i$} &{Total required bit-rate of client $i$} \\
\hline
\centering{$\eta_j$} &{The technology efficiency of WNP $j$} \\
\hline
\centering{$\vartheta_j $} &{Allocated spectrum to WNP $j$} \\
\hline
\centering{$L_j^f $} &{The load of WNP $j$ measured as sum of all assigned bit-rates to clients at price controlling cycle $f$} \\
\hline
\centering{$L_j^{max} $} &{Maximum bit-rate capacity of WNP $j$ under current spectrum allocation $(L_j^{max}=\eta_j\vartheta_j)$} \\
\hline
\centering{$L_j^{A}(t) $} &{Available bit-rate capacity of WNP $j$ at iteration $t$ in each price controlling cycle} \\
\hline
\end{tabular}
\end{table}

\section{Problem Formulation}
\subsection{The System Model}
In this paper, we propose a solution for fair pricing in HWAN at the presence of a regulatory agent as the moderator. Considered wireless networks are heterogeneous on their used technology, coverage area, or ownership (WNP).
Theoretically the considered area and the coverage area of each WNP do not make any changes on the problem, and the proposed solution with inclusion of heterogeneity in different forms is one of the key contributions of the work. 
For a better illustration, we can consider a smart city scenario where different types of users, sensors, cameras, and actuators exist with different requirement in a large geographical area. All clients are assumed to be low mobility. The considered area is covered by multiple WNPs, while each WNP may cover all or a part of the area. Here each client requests data from different covering WNPs based on their price and quality. 
%
%Figure \ref{fig:1} shows three different time scales of fair pricing problem.

\begin{figure}
\begin{center}
\includegraphics[width=.6\columnwidth]{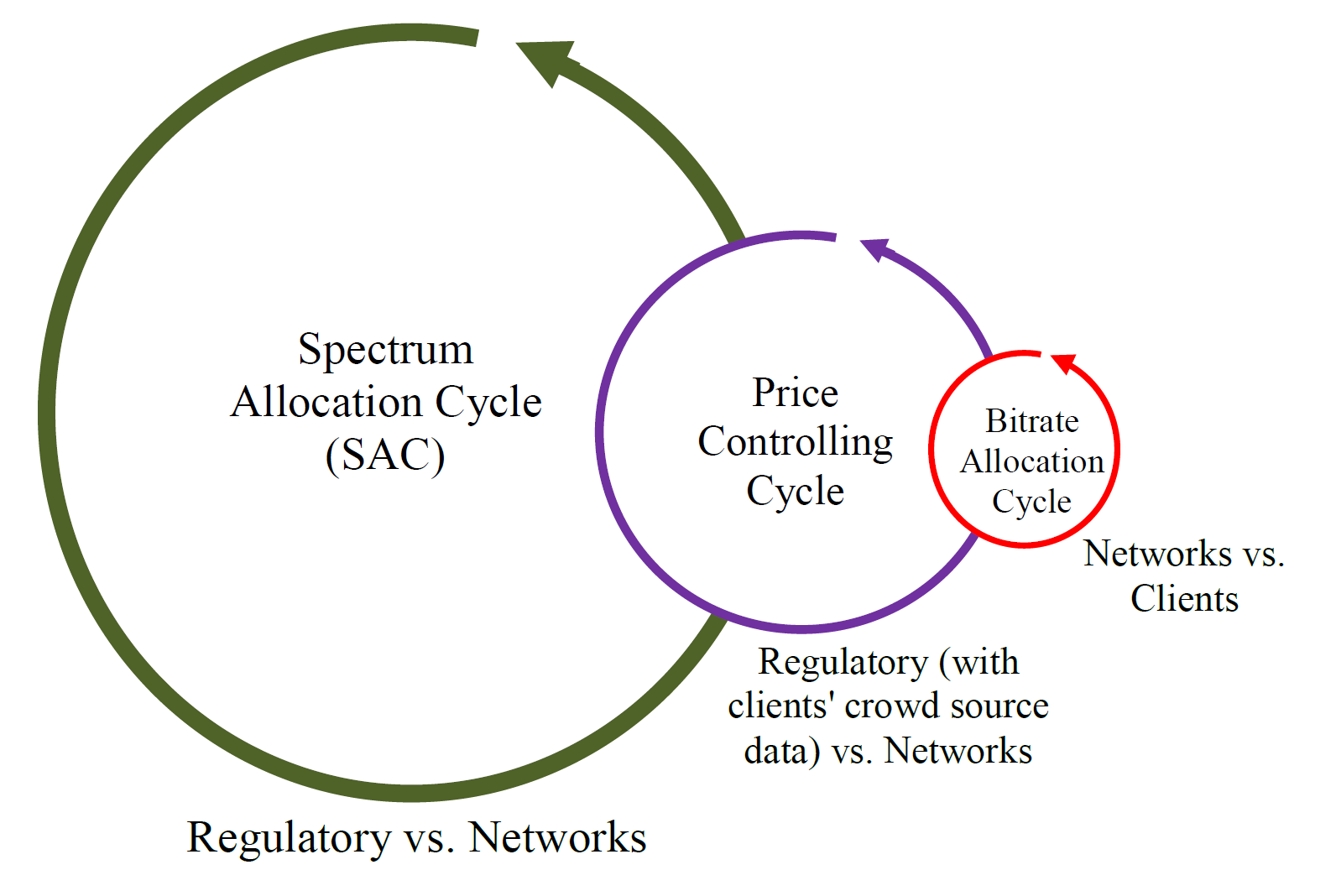}
\end{center}
\caption{\footnotesize Time scales for decisions}\label{fig:1}
\end{figure}

The chain of pricing and bit-rate allocation, involves  three decision problems. These problems are solved by different agents at different time-scales. 
As shown in Figure \ref{fig:1}, spectrum allocation is applied by the regulator in long time periods. Price controlling is a trading between WNPs and regulator to announce a fair price in medium time periods, and finally preparing the request bundles by clients and acceptance of them by WNPs are done in short time periods.
This paper proposes a fair price control system based on implicit feedback received form the clients in the form of their preferred service demands that triggers the regulator to impose a cap on service pricing on network specific basis. Figure \ref{fig:1} shows that in each Spectrum Allocation Cycle (SAC), there are Price Controlling Cycles (PCCs) and in each PCC, there are Bit rate Allocation Cycles (BACs). The proposed mechanism cannot be modeled as a Stackelberg game since the stages have different time scales and they have multi-level interactions which will be explained in next Sections. 

The idea used in the proposed mechanism is deduced from real world where we do not explicitly know the real and fair price for a commodity but we can evaluate the relative merit of the commodities and can make judgements on fairness of prices of the commodities offered by different sellers. Since the clients are the receiving end of the services, their judgement on the price fairness based on their personal cost versus quality trade off would be the best reference by the market moderator to discover whether the pricing of a WNP is excessive or fair. This triggers the regulator to impose a cap on service pricings on per WNP specific basis. The price adjustment is performed in cycles to respond to changes in system dynamics. The proposed model can be easily implemented in real world and especially IoT scenarios. It requires more dynamic  payment mechanisms at implementation level which can be done via Blockchain-based smart contracts \cite{Pascale_smartContract20,umoren2020blockchain}

The desired equilibrium is finding a solution in which prices are fair, and  the clients' service requests are satisfied. In this system the fair service price is considered to be just above the marginal costs of the a WNP. The proposed scheme can be used in competitive and non-competitive environments alike in the context of WNPs, CRNs and HAWN in general. It does not require any prior knowledge of WNPs’ marginal cost nor clients’ preferences and is robust and at the same time resilient against selfish behavior of the WNPs while guaranteeing their healthy profit margins. 

As presented in Figure \ref{fig:3} the regulator receives the perfect request bundles ($\{\boldsymbol{s^i}\}$) from clients, current load of networks ($\boldsymbol{L}$) and current announced prices ($\boldsymbol{p}$); by monitoring the market, it decides on the maximum allowed price for each WNP ($\tilde{p}_j$). The term ``perfect request bundle" (PRB) refers to the ideal service bundle of a client without considering WNPs’ service capacity constraints and unfair prices. The clients rank the WNPs based on their preferences and prepare their PRBs according to the rank of the WNPs and their current service prices. A client may potentially request services from multiple WNPs. Due to WNPs’ resource constraints, the actual service delivery to a client may be different from its preferred service demand. Using PRB information received from the clients, our proposed mechanism faithfully discovers the fair service price, just above the marginal costs of the WNPs.

The Table \ref{Tbl:1} provides the list of variable notations used in this paper.

% \begin{figure}
% \begin{center}
% \includegraphics[width=.97\columnwidth]{pics/Fig002.png}
% \end{center}
% \caption{\footnotesize HWAN environment}\label{fig:2}
% \end{figure}

\subsection{Game-theoretic problem formulation}
%As summarized in Figure \ref{fig:3}, we propose a three-tire game model for fair pricing and bit-rate allocation. %In the rest of the paper we may used '\textit{network}' that means WNP. 
In the proposed mechanism, clients prepare their service demands on a selfish manner to maximize their utilities, and they weigh the WNPs according to networks’ state vector and their service pricings. The network state vector is the current profile of a wireless network that involves: delay, jitter, power strength, security level, current load, and total capacity. The network state vector is considered as a global knowledge that all clients can access. 

WNPs know that the effect of service pricing on the clients’ weighting, and the regulator assigns the maximum selling price for each network according to clients’ perfect request bundles. 
%Perfect request bundles are crowd sourced from clients. The regulatory guides the market to the state where the prices are fair; so, users and networks all get the appropriate desirability.

The proposed game consists of three stages where in the first stage, the regulator adjusts the maximum selling price based on PRBs that are crowdsourced from clients and final allocated bit-rates from networks to clients. WNPs announce the price of their services in the second stage based on the predefined maximum selling price and their marginal costs. In the third stage clients weigh the networks, prepare the crowdsource data for the regulator, and the request bundle to buy the service from networks.

\begin{figure}
\begin{center}
\includegraphics[width=.7\columnwidth]{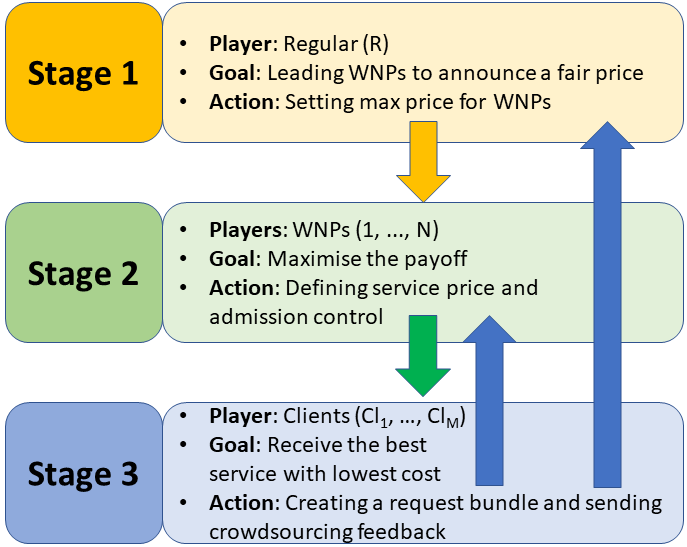}
\end{center}
\caption{\footnotesize Three stage of the CSPC for price controlling and bit-rate allocation.}\label{fig:3}
\end{figure}

%The extensive form of the proposed game is shown in Figure \ref{fig:3B}. 
Considering a dynamic game tree the regulator is located in the root of the game tree (first level). The action of the regulator is determination of maximum allowed prices ($\tilde{p}^{j}$). WNPs are in the second level and clients are in the third level. Their actions are shown in Figure \ref{fig:3}. From the sub-game-perfect Nash equilibrium theorem, we know a strategy profile is a subgame perfect equilibrium if it represents a Nash equilibrium of every subgame of the original game \cite{mackenzie2006game}. To solve the proposed game by using backward induction approach, first, the best response of clients $(\boldsymbol{s^{i*}},\boldsymbol{r^{i*}})$ in the third level is calculated, which is the Nash equilibrium of third level according to the decisions of agents in previous levels. Then the best response of WNPs in second level (the price of service ($p^{j*}$) is calculated which depends on the load and defined maximum allowed price by the regulator. Finally the best decision of the regulator to adjust the maximum allowed prices ($\tilde{p}^{j*}$) is calculated.    

%\begin{figure}
%\begin{center}
%\includegraphics[width=.8\columnwidth]{pics/Fig003B.png}
%\end{center}
%\caption{\footnotesize Extensive form of proposed three stage game.}\label{fig:3B}
%\end{figure}
%
\section{Crowdsourcing Price Control (CSPC) mechanism}
In this section, we describe our proposed CSPC mechanism starting with client strategy in the bottom stage and moving up to the regulator strategy at the top.
\subsection{Clients' Strategies}
The strategy of the clients involves the three main actions:
\subsubsection{WNP weighting}\label{Subsec:weight}
We consider that the client have an initial weight for each WNP, 
and the clients adjust the weights of WNPs based on their degree of honesty in pricing. The WNPs are heterogeneous in terms of their ownership, technology efficiency, and coverage.

Suppose $\bar{P}$ be the average of all announced prices of all WNPs in the market $(\bar{P}=\frac{1}{N} \sum_{j=1}^N p_j)$; note that the price announce by each WNP is uniform (same for all its clients). Let $\boldsymbol{\rho}^i=(\rho_1^i,\dots,\rho_N^i)$ be the suitable ratio of prices (SRP) from point of view of client $i$, that $\rho_j^i$ is the fair ratio of the price of WNP $j$ to the average of all WNP prices ($\bar{P}$). Hence the fair price estimation of client $i$ is $\boldsymbol{\hat{p}^i}=\bar{P}.\boldsymbol{\rho^i}$ which is the suitable price from the point of view of client. Base on the nature of Marginal Cost for preparing the service for WNPs, the SRP of clients are approximately near the marginal costs ratio that is proved as a continuation; $\boldsymbol{\rho}^i\approx(\frac{MC_1}{\bar{MC}},\dots,\frac{MC_j}{\bar{MC}} ,\dots,\frac{MC_N}{\bar{MC}})$ where ($\bar{MC}$) is the average of all marginal costs ($\bar{MC}=\frac{1}{N}\sum_{j=1}^N MC_j$). It is must be mentioned that the marginal cost of WNP $j$ ($MC_j$) is the network's private information.
Let $\boldsymbol{\tilde{w}}^i=(\tilde{w}_1^i,\dots,\tilde{w}_N^i )$ show the initial weight client $i$ for $N$ available WNPs and $\boldsymbol{p}=(p_1,\dots,p_N)$ is the announced price vector. Clients adjust their weight vectors according to their initial belief of weights and the announced prices for services. If the announced price for the service of a WNP was higher than the client's estimation, the weight of the WNP will be reduced by the client and if the announced price for the service of a WNP was lower than the client's approximation, the weight of the WNP would be increased. We use the following relation to simulate the weight adjustment:
\begin{equation}\label{Eq_1}
%\begin{array}{l}
   w_j^i=\tilde{w}_j^i (1+ \beta  \frac{\hat{p}_j^i-p_j}{p_j})=\tilde{w}_j^i (1+ \beta  \frac{\hat{p}_j^i}{p_j}-\beta),
%\end{array}
\end{equation}
where $\hat{p}_j^i$ is the estimated fair price of WNP $j$ by client $i$ %\textcolor{green}{
and $\beta(>0)$ is the weight adjustment coefficient that controls the rate of increase or decrease of a WNP's weight according to the portion of estimated fair price and real price of a WNP.
It is believed that if $\hat{p}_j^i>p_j$ then $w_j^i>\tilde{w}_j^i$ that means the weight of WNP is increased and if $\hat{p}_j^i<p_j$ then $w_j^i<\tilde{w}_j^i$ that means the weight of WNP is decreased for client $i$. If all WNPs are honest, they announce the price according to their real marginal cost ($\forall j\in\{1,\dots,N\}: p_j=MC_j (L_j )$). Since the marginal cost is a function of the WNP's load ($L_j$), then the clients' price approximation is near/equal to the announced price $(\boldsymbol{\hat{p}}^i\approx\boldsymbol{p})$ and the WNPs' weight are the same as their initial weights $(\boldsymbol{w}^i=\boldsymbol{\tilde{w}}^i)$, as shown below:
%
%\begin{equation}\label{Eq_2}
\begin{align}\label{Eq_2}
   \boldsymbol{\hat{p}}^i&=\bar{P}\cdot\boldsymbol{\rho}^i, \\
   \boldsymbol{\hat{p}}^i &\approx 
   \bar{P}\cdot(\frac{MC_1}{\bar{MC}},\dots,\frac{MC_N}{\bar{MC}}), \nonumber\\
    &\approx 
   \frac{1}{N} \sum_{j=1}^N p_j(\frac{MC_1}{\bar{MC}} ,\dots,\frac{MC_N} 
{\bar{MC}}),\nonumber\\ %(p_j=MC_j(L_j)) 
   & \approx
\frac{1}{N} \sum_{j=1}^N MC_j(\frac{p_1}{\bar{MC}} ,\dots,\frac{p_N}
{\bar{MC}}),\nonumber\\
 & \approx
\bar{MC}\cdot(\frac{p_1}{\bar{MC}} ,\dots,\frac{p_N}
{\bar{MC}}),\nonumber\\
 & \approx (p_1,\dots,p_N)\nonumber
 & \approx \boldsymbol{p}.\nonumber
\end{align}
%\end{equation}
%
Knowing $(p_j=MC_j(L_j))$ and using (\ref{Eq_1}), we can write $w_j^i \approx \tilde{w}_j^i$.
Now, suppose one WNP (e.g. WNP $k$) announces its price higher than its marginal cost $(p_k=MC_k+\Delta)$ and other WNPs are being honest announcing a price the same as their marginal cost $(p_{-k}=MC_{-k})$, hence in this situation $\bar{P}=\frac{1}{N} (\sum_{j=1}^N MC_j +\Delta)$, so the weight of WNP $k$ for client $i$ will be:
\begin{equation}\label{Eq_3}
w_k^i=\tilde{w}_k^i (1+ \beta  \frac{\bar{P} \rho_k^i}{MC_k+\Delta}-\beta),
%w_k^i=\tilde{w}_k^i (2\frac{\bar{P} \rho_k^i}{(MC_k+\Delta)}-1),
\end{equation}

while for the honest WNPs $(j\in-k)$, the weights are:
\begin{equation}\label{Eq_4}
w_j^i=\tilde{w}_j^i (1+ \beta  \frac{\bar{P} \rho_j^i}{MC_j}-\beta).
%w_j^i=\tilde{w}_j^i (2\frac{\bar{P} \rho_j^i}{MC_j}-1).
%\end{array}
\end{equation}
Using the relations \eqref{Eq_3} and \eqref{Eq_4}, we conclude that the weight of an unfair WNP decreases, and weights of fair WNPs have an increase because $\bar{P}$ is increased by unfair pricing of WNP $k$. Hence, when a WNP, suggests an unfair price, the clients decrease its weight. The reduction of the weight of a WNP is the same as clients' defined penalty for the WNP and it results in the reduction of the requested amount of service. 

\subsubsection{Preparing Crowdsourcing Information}
After WNP weighting, clients decide on the amount of bit-rate received from each one of available WNPs. The utility of client $i$ is expressed in the form of following Constant Elasticity of Substitution (CES) utility function \cite{Ref_8}. 
\begin{equation}\label{Eq_5}
u^i (\boldsymbol{x}^i,\boldsymbol{w}^i )=\left(\sum_{j=1}^N \sqrt[r]{w_j^i x_j^i}\right)^r,
\end{equation}
where $x_j^i$, is the total amount of bit-rate assign to client $i$ from WNP $j$ in current PCC. The CES utility functions are increasing and concave for $r>1$ that is widely used for optimization problems.

Suppose $p=(p_1,\dots,p_N ) \gg 0$ is the price vector of $N$ available WNPs in current PCC, and $R^i$ is the total required bit-rate and $H^i$ is the total budget of client $i$.
Clients prepare the crowdsource data using the following optimization problem:
\begin{subequations}\label{Eq_6}
\begin{align}
\max_{s^i_j} &\thickspace u^i\left(\boldsymbol{s^i},\boldsymbol{w}^i\right) \\
\text{Subject to:     } %&\nonumber\\
& \boldsymbol{\hat{p}^i} \cdot \boldsymbol{s^i} \leqslant H^i\\
&\sum_{j=1}^N s_j^i \leqslant R^i \\
& s^i \in \mathbb{R}_+^N.
\end{align}
\end{subequations}
The first constraint is the budget constraint. The second constraint, controls the total requested bit-rate do not exceed from the total required bit-rate and the third constraint avoids the requested amounts to be negative. Each client sends the perfect request bundle $\boldsymbol{s^i}=(s_1^i,\dots,s_N^i)$ as the crowdsourcing data to the regulator.

\subsubsection{Preparing the Request bundle}
Due to load constraint for each WNP, the bit-rate allocation is a repeating process within each PCC that is called bit-rate Allocation Iteration (BAI). As the price of service changes, it causes the clients to change their request bundles to fulfill their required service and maximize their utilities. Hence, the bit-rate allocation to clients is an iterative mechanism. In each PCC, there are BAIs as shown in Figure \ref{fig:1}.  Let $x^i (t)=(x_1^i (t),\dots,x_N^i (t))$ determine the assigned bit-rate to client $i$ by WNP $j$ at $t^{th}$ BAI in current PCC. 

Suppose $\boldsymbol{r^i} (t)=(r_1^i(t),\dots,r_N^i(t))$ is the request bundle prepared by client $i$ at $t^{th}$ BAI. Each client tries to increase its payoff trough request bundle preparing step that is mathematically shown as follow:
\begin{subequations}\label{Eq_7}
\begin{align}
\max_{r^i} &\thickspace u^i\left(\boldsymbol{r^i}(t),\boldsymbol{w}^i\right) \\
\text{Subject to:    } 
&\boldsymbol{p} \cdot \boldsymbol{r^i}(t)\leqslant H^i(t) \label{7:1}\\
& \sum_{j=1}^n r_j^i \leqslant R^i(t) \label{7:2}\\
& r_j^i(t) \leqslant L_j^A(t) \label{7:3}\\
& \boldsymbol{r^i}(t) \in \mathbb{R}_+^n. \label{7:4}
\end{align}
\end{subequations}
Here $L_j^A (t)$  is the available bit-rate from WNP $j$; $L_j^A (t) =L_j^{max}-\sum_{\tau=1}^{t-1} \sum_{i=1}^M x_j^i (\tau)$ and $L_j^{max}=\vartheta_j \eta_j$. Also $\vartheta_j$ is the allocated spectrum to WNP $j$ and $\eta_j$ is its technology efficiency. Total allocated bit-rate to client $i$ until turn $t$ of current PCC is $\sum_{\tau=1}^{t-1} \sum_{j=1}^N x_j^i (\tau)$, and its remained request is $R^i (t)=R^i-\sum_{\tau=1}^{t-1} \sum_{j=1}^N x_j^i (\tau)$ , while its remained budget is $H^i (t)=H^i-\sum_{\tau=1}^{t-1} \sum_{j=1}^N p_j x_j^i (\tau)$.
The first constraint in (\ref{Eq_7}) essentially demonstrates the client’s budget constraint and confines the client to buy the bit-rate from different WNPs more than its budget. The second constraint is the client requirement constraint that controls the received bit-rate from all networks. The third constraint is the WNP load constraint, which controls the preparing the request from a WNP more than its available load. 
The solution of problem (\ref{Eq_7}) for client $i$ is $r^i (t)$, that is the client’s request bundle in iteration $t$, which relies on price vector and the assigned spectrum to WNPs.

\begin{Definition}\label{Def_1}
Clients' Request bundle:
\end{Definition}
The result of \eqref{Eq_7} shows the client reaction as function on its budget $(H^i)$, its required bit-rate $(R^i)$, its preferences $(w_j^i)$ and available bit-rates from WNPs $\{L_j^A;j=1,\dots,N\}$. In other words  $r^i (t)=(r_1^i(t),\dots,r_N^i(t))$ is a function of $(H^i (t),R^i (t),w^i,\{L_j^A (t)\})$.

\subsection{WNP Strategy}
Each WNP strategy consists of the following actions:
\subsubsection{Price announcement}
According to proposed mechanism, it is not profitable for WNPs to announce a price, higher than their real marginal cost, which is achieved from the WNP's cost and suitable payoff. However, we have to answer a basic question \emph{``What is a suitable price for the prepared service of a WNP?"}.

From the studies in Economics we know that 
the marginal cost (MC) curve is U--shaped and total cost curve is a curve with a turning point \cite{Ref_2}. The marginal cost curve is the derivative of the total cost (TC) curve $MC(l)=\frac{\partial}
{\partial l}TC(l)$. The MC curve shows the increase of cost against the increase a unit of produced service. The microeconomic theory says that in the stationary state of market, the MC of the product is equal to the market suitable price \cite{Ref_2}. For example, %as shown in Figure \ref{fig:4}
suppose for a specific WNP the price of service be $10$, hence, the suitable amount of service $(\tilde{L})$ is a point where the difference of WNP's income and TC is maximized. The slope of TC in that point is equal to the slope of total income (TI) (i.e. $\frac{\partial}
{\partial l}TC(\tilde{L})=\frac{\partial}
{\partial l} TI(\tilde{L})=10$). We know the slope of the TI is equal to the service price ($p=10$ in our example), hence we can find the suitable amount of product is $\Delta$ using $MC (MC(\tilde{L}=\Delta)=10)$.
In microeconomic it is proved that if the price of a product be less than the company marginal cost, producing such product is not profitable for the company and while the price of product be equal or higher than its marginal cost, the product is profitable \cite{Ref_8}.

Suppose $\vartheta_j$ is the current assigned spectrum to WNP $j$. The WNP hopes to sell all possible throughput $L_j^{max}=\vartheta_j \eta_j$. So, as described in Subsection \ref{Subsec:weight}, if a WNP announces a higher price more than its marginal cost, it loses the client satisfaction leading to the loss of weight.

The regulator crowdsources the perfect request bundles and using the proposed solution in the next section, it finds which WNPs are unfair and which ones are honest. So, the regulator forces the unfair WNP to reduce the price of the service. In long term, the regulator can reduce the assigned spectrum to unfair WNPs which is not within the scope of this work. We call a WNP ``an honest WNP" if it announces its price at most equal to its marginal cost or the maximum allowed price by the regulator. Against the definition of honest WNP, we call a WNP that announces the maximum allowed price for pricing as an ``unfair WNP". Hence, the pricing strategy is:
%$\max{(MC_j(\vartheta_j \eta_j),\tilde{p}_j)}$
%
\begin{equation}\label{Eq_8}
p_j = \left\{
\begin{array}{rl}
\min \big(MC_j(\vartheta_j \eta_j),\tilde{p}_j \big)& \text{if } \text{WNP}_j \text{ is honest}\\
\tilde{p}_j & \text{if } \text{WNP}_j \text{ is unfair}
\end{array} \right.
\end{equation}
where $\tilde{p}_j$ is the price ceiling or maximum allowed price for a unit of the service of WNP $j$. Of course, as it will be stated in the following section, the regulator strategy adjusts the price ceiling vector $(\boldsymbol{\tilde{p}})$ so that $\tilde{p}_j\rightarrow MC_j(\vartheta_j \eta_j)$. 

\subsubsection{bit-rate allocation}
WNPs' strategy for bit-rate distribution is accept requests until the load constraint allows. Suppose $\boldsymbol{r^i} (t)=(r_1^i,\dots,r_N^i)_t$ is the request bundle prepared by client $i$ and $L_j^A (t)$ is the available bit-rate from WNP $j$ in round $t$. The network decision for accept or reject the requests is as the following:
\begin{equation}\label{Eq_9}
x_j^i (t) = \left\{
\begin{array}{rl}
r_j^i & \sum_{i=1}^m r_j^i \leqslant L_j^A(t)\\
lottery-based & \text{otherwise},
\end{array} \right.
\end{equation}
in lottery-based, when available bit-rate is less than total requests we find the index $k$ where the required bit-rates could be fully covered ($\sum_{i=1}^k r_j^i \leqslant L_j^A(t)$).
\begin{equation}\label{Eq_10}
x_j^i (t) = \left\{
\begin{array}{rl}
r_j^i  & i \leqslant k \\
L_j^A (t) -\sum_{l=1}^k r_j^l  & i=k+1\\
0 & i > k+1.
\end{array} \right.
\end{equation}
%
%\\\\
%a. If $\sum_{i=1}^m r_j^i \leqslant L_j^A(t)$
%\begin{equation}\label{Eq_9A}
%x_j^i (t)=r_j^i,
%\end{equation}
%b. Else\\
%-\qquad For $i$=1 to $m$\\
%-\qquad BEGIN\\
%-\qquad\qquad IF $L_j^A(t) \leqslant 0$ THEN\\
%-\qquad\qquad\qquad BREAK;\\
%-\qquad\qquad ELSE\\
%-\qquad\qquad BEGIN\\
%\begin{equation}\label{Eq_10A}
%\begin{array}{l}
%x_j^i (t)=\min( L_j^A (t),r_j^i);\\
%L_j^A (t)=L_j^A (t)-x_j^i (t)
%\end{array}
%\end{equation}
%-\qquad\qquad END\\
%-\qquad END\\
%
Finally, for each WNP $j$, all requests will be accepted or $L_j^A (t)$ meet zero and part of requests are accepted. The utility of WNP is defined as its total revenue.
\begin{equation}\label{Eq_11}
\pi_j=p_jL_j-TC(L_j^{max}),
\end{equation}
where $p_j L_j$ is the total income and $TC(L_j^{max} )$ is the total cost for preparing the amount of $L_j^{max}$ service.

\subsection{Regulator strategy}
The spectrum adjustment is applied in long term time intervals, but the price controlling is done in middle time intervals. The policy of regulator for control the prices is setting a price ceiling for each WNP.
Suppose $\boldsymbol{\vartheta}=(\vartheta_1,\dots,\vartheta_N)$ be the current Spectrum Allocation Table (SAT). The regulator crowdsources the perfect required bit-rates of clients as described in (\ref{Eq_6}) and monitors the current load of WNPs. According to gathered data three different conditions occur that are shown in Table \ref{Tbl:2}. Suppose $S$ defines perfect required service from WNPs that is aimed from the crowdsourced data where $S_j=\sum_{i=1}^m s_j^i $.

\begin{table}
\caption{Different situations for each PCC}\label{Tbl:2}
%\vspace{4pt} \noindent
\begin{tabular}{|p{.08\columnwidth}|p{.25\columnwidth}|p{.5\columnwidth}|}
\hline
{ID} & {Condition} &{Analysis}\\
\hline
{1} & {$L_j<S_j$ and  $L_j<L_j^{max}$} &{The announced price by WNP $j$ is more than fair price}\\
\hline
{2} & {$L_j<S_j$ and  $L_j=L_j^{max}$} &{Assigned spectrum to network $j$ is less than the market request}\\
\hline
{3} & {$L_j>S_j$} &{WNP $j$ announces fair price for its service}\\
\hline
\end{tabular}
\end{table}

Suppose $\boldsymbol{\tilde{p}}=(\tilde{p}_1,\dots,\tilde{p}_N )$ be the ceiling price vector. \eqref{Eq_12} describes the proposed strategy for adjustment of ceiling price for WNPs: 
\begin{equation}\label{Eq_12}
\tilde{p}_j = \left\{
\begin{array}{rl}
p_j \frac{L_j}{S_j} \xi & \text{if } L_j \geqslant S_j \\
p_j \max{(\frac{L_j}{S_j},\gamma)} & \text{otherwise}
\end{array} \right.
\end{equation}
%\textcolor{green}{
In this relation, the element $ \frac{L_j}{S_j}$ controls the rate of increase of maximum allowed price, hence, we consider it as the adjustment rate parameter, that satisfies the bellow goals:
%}
\begin{itemize}
	\item WNPs that have higher load than perfect request, are allowed to increase their price;
	\item WNPs that have lower load than perfect request, are enforced to decrease their price;
	\item As the difference of $L_j$ and $S_j$ be higher, the WNP reward or punishment is higher;
	\item In steady state when $S_j\thickapprox L_j$, we meet that $\frac{L_j}{S_j}$  is very low, that results in the changes on $\tilde{p}_j$s be very low. 
\end{itemize}

If a constant value for adjustment rate parameter was used in \eqref{Eq_12} we would find that in steady state where $S_j\thickapprox L_j$, the system diverges; suppose an unfair WNP that regulator could control its price by enforcing rule. At this time, clients suggest the unfair WNP as an honest WNP $(L_j \geqslant S_j)$ because its price is very near to fair price. By constant value for adjustment rate suddenly the unfair WNP is allowed to increase its price. This event causes the instability in system convergence. However, as explained above using the $\frac{L_j}{S_j}$ for adjustment rate parameter the proposed rule in \eqref{Eq_12} does not diverge the $\tilde{p}_j$s from steady state.

Hence, unfair WNPs are punished and they are enforced to decrease their announced price. Honest WNPs are allowed to increase their price as much as their adjustment rate parameter $\frac{L_j}{S_j}$ allows. So, the coefficient $\xi(>1)$  in (\ref{Eq_12}) is additional regulator reward for honest WNPs. The value of parameter $\xi$ plays an important role in system convergence. As $\xi \rightarrow 1$, we see that system needs extra iterations to converge because the initial values for $\tilde{p}$ are initiated randomly and they are far from real marginal costs, all prices are less than marginal costs, hence the opportunity for WNPs to increase their price, at least to their marginal cost is very low. On the other hand, for $\xi>2$ the system reward for honest detected WNPs is very high and causes instability. Suppose an unfair WNP is enforced to decrease its price, when our algorithm is in its convergence iterations, and the price of unfair WNP is going to its marginal cost, suddenly we reward the unfair WNP and allow it to double its price. Therefore, we find that $\xi=1.05$ is suitable value for this parameter.
The parameter $\gamma \in (0,1)$ is the minimum penalty coefficient for reduction of price of unfair network in (\ref{Eq_12}). For example, when $\gamma=0.9$, it guarantees that the punishment of unfair WNPs be at most 10\% decrease in price in each PCC. Low values of $\gamma$ causes the instability of system and by detection of any small cheat of a WNP, its price is reduced highly and very high value of $\gamma$ cause to unfair WNPs not be punished.

\begin{figure}
\begin{center}
\includegraphics[width=.75\columnwidth]{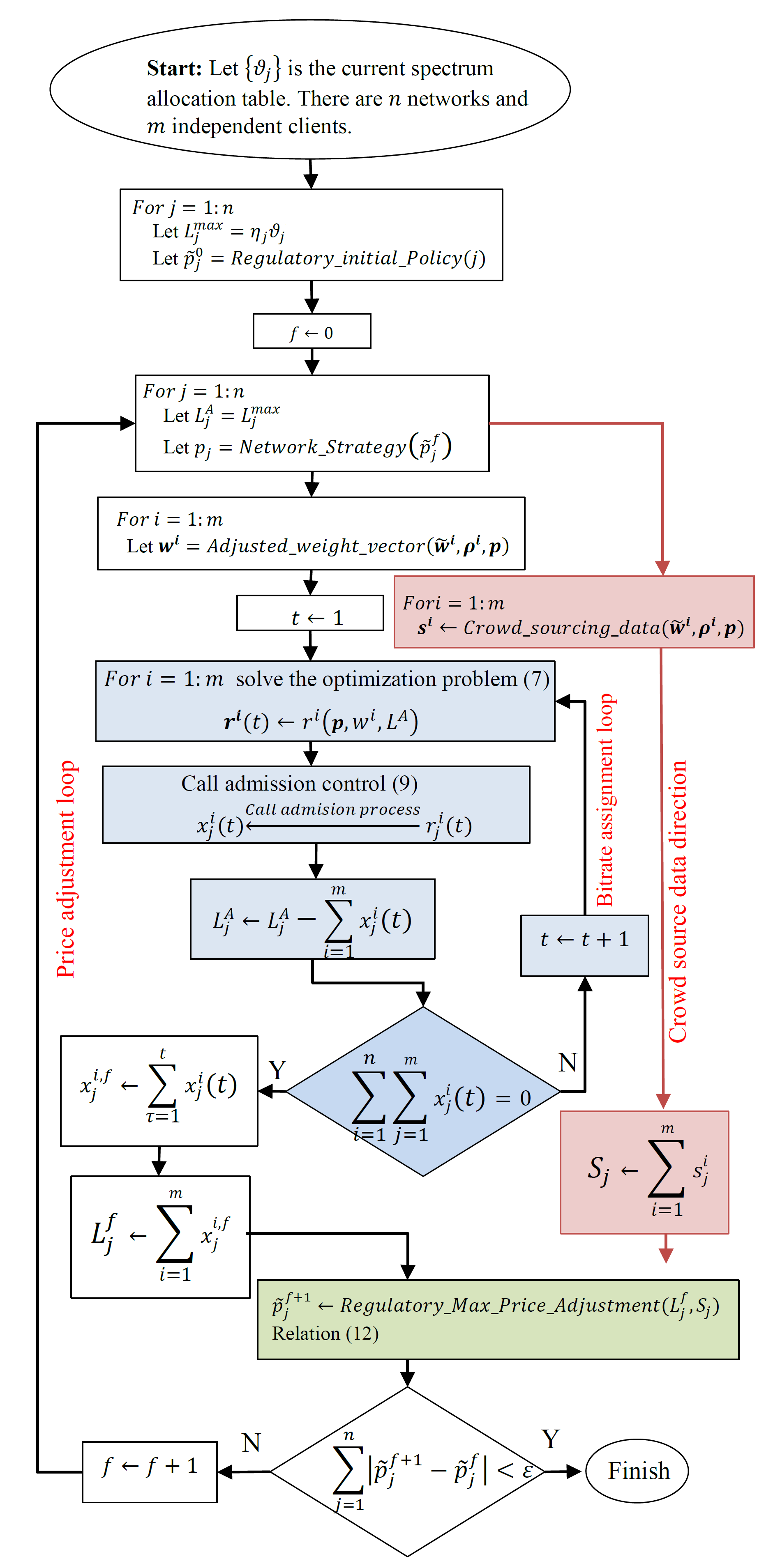}
\end{center}
\caption{\footnotesize Flowchart of CSPC approach for resource allocation and fair pricing.}\label{fig:5}
\end{figure}

\subsection{Summarizing CSPC mechanism}
In Figure \ref{fig:5}, the overall process of CSPC, is shown. The process contains of two loop, in the internal loop, the bit-rate allocations are calculated for clients, and in the outer loop, the maximum allowed price for the service of WNPs is adjusted until the prices meet their optimum value. 

To compute the time complexity of proposed mechanism we should indicate the independent variables in our model. Let us denote the time complexity of solving the best response problem of client $i$ as $g_i$ (the optimisation problem) which is related to number of constraints and WNPs. Then $g_i$ becomes one of independent variables that we consider. The other independent variables are the number of WNPs ($n$), number of clients ($m$), maximum number of iterations to assign the bit rates to clients ($T$), and the Maximum Rounds for Convergence of Prices ($F$). Hence, The Time complexity of our approach will be $O(\sum_{i=1}^mFTg_i)$. It should mentioned that all clients solve the optimization problem simultaneously; so, the coefficient $m$ in the relation is eliminated and the order of time complexity becomes $O(FT\hat{g})$ where $\hat{g}=max\{g_i\}$.

\section{Analysis of the model}
\subsection{The consistency of the WNP's marginal cost and clients' demand function}
In this paper, we focused on crowdsourced data gathering and decision. The main idea of this paper is based on the rationality of the clients and WNPs in HWAN. We considered that clients are relatively aware of service value. A user may mistake the service value, but when a number of users attend the valuation, the results are very close to reality. A WNP also knows its clients' taste for service meaning that the WNP knows that ``how much money, clients are ready to spend to buy its prepared service" and also we should add the basic cyclic rules of rationality as ``The Clients are rational", ``WNPs are rational and know that clients are rational", ``Clients are rational and know WNPs are rational and they know that clients are rational" and the principle of rational decision making leads to a hypothesis in the mind that the real value of the service provided by the WNPs is proportional to the purchasing power of users.

{\textbf{Lemma 1)}} \textsl{The marginal cost curve for a rational WNP is always prepared according to clients demand function.} \\
In a HWAN, rational WNP $j$ with Marginal Cost $MC_j (L_j )$, stays in the market if clients' demand for its service with price $p_j$  be equal or more than $MC_j^{-1} (p_j )$. 
\begin{equation}\label{Eq_13}
MC_j \left(Dm_j (p_j,p_{-j})\right) \leqslant p_j,
\end{equation}
where $Dm_j (\cdot)$ is the clients' demand function for WNP $j$.\\
\textbf{Proof:}\\
Suppose for a specific price $p_j$, the clients' demand be less than the prepared service, hence it is not profitable for the WNP to continue its market activity and it leaves the market. We know that the WNP is rational and it adjusts its costs to achieve suitable and profitable marginal cost. Hence, always the cost of preparing the service is followed by clients belief.   
In other words, for $p_j$, if the clients' demand $Dm_j (p_j,p_{-j})$ be less than the prepared service $MC_j^{-1} (p_j)$, the WNP adjusts its costs to achieve new marginal cost $\acute{MC}$ where $Dm_j (p_j,p_{-j})=\acute{MC}_j^{-1} (p_j )$.\\
\qed\\
According to Lemma 1, pricing for the prepared service, according to the marginal cost which is aimed from the inception of the market, is always profitable for the WNP.

\subsection{Simulation setup}
We defined two settings to generate data by simulation to use for different scenarios evaluating the proposed CSPC schema. In simulation setting 1, we have $50$ clients and $3$ WNPs with the spectrum allocation table $\vartheta=[30, 48, 60]$ (MHz) and in setting 2, we have $100$ clients and $6$ WNPs and the spectrum allocation table is $\vartheta=[30,48, 60, 49, 75, 27]$ (MHz).
%Description:Explain about design of two datasets.
In simulation setting 1, the regulator assigns low, middle and high amount of spectrum to the three different WNPs, respectively. According to the marginal costs of preparing the service, the prepared service of each WNP has different price that in simulations we look for the ability of proposed algorithm to find the marginal cost using crowdsourced information. For marginal costs we considered \$ as a nominal currency and the prices are relative and may not match actual value in \$. In simulation setting 2, the number of networks are increased and we are looking for the scalability of our proposed algorithm. 
Other parameters of these settings are shown in Table \ref{Tbl:3}. The rationale behind these settings is to cover a variety of potential situations, including different technology efficiency levels $\eta_j$, different assigned spectrum, $\vartheta_j$, and marginal cost.

Our simulation are implemented in MATLAB 2014 environment and run on a computer with Windows 10 (64bit), Core i5, 8GB RAM. The required time for simulation setting 1 that has three WNPs and 50 clients was about two hours and the required time for simulation setting 2 with 6 WNPs and 100 client is about 3 hours. We used {\it{parfor}} command to simulate concurrency of decisions.

\begin{table}
\begin{center}
\caption{Definition of simulation parameters for defined settings.}\label{Tbl:3}
%\vspace{4pt} \noindent
\begin{tabular}{|p{.3\columnwidth}|p{.3\columnwidth}|p{.3\columnwidth}|}
\hline
%\footnotesize
{}&{\centering \textbf{Simulation Setting 1}} & {\centering \textbf{Simulation Setting 2}}\\
\hline
{Number of clients ($M$)}&{50} & {100}\\
\hline
{Number of networks ($N$)}&{3} & {6}\\
\hline
{Clients estimation ratio ($\boldsymbol{\rho^i}$)}&{$\rho_j^i=(1 \pm 0.1)$ (10\% tolerance)} & {$\rho_j^i=(1 \pm 0.1)$ (10\% tolerance)}\\
\hline
{Allocated spectrum (MHz)}&{[30 48 60]} & {[30,48, 60, 49, 75, 27]}\\
\hline
{Efficiency ($\boldsymbol{\eta}$) (bps/Hz)}&{[8, 9, 6]} & {[8, 9, 6, 5 , 4, 7]}\\
\hline
{Prepared Service (Mbps)}&{[240, 432, 360]} & {[240, 432, 360, 245, 300, 189]}\\
\hline
{Marginal Cost of WNPs (\$)}&{[19.68, 38.68, 28.73]} & {[19.68, 38.68, 28.73, 9.79, 14.18, 6.97]}\\
\hline
{$\xi$ , $\gamma$ , $\beta$}&{1.05, 0.9, 2} &{1.05, 0.9, 2}\\
\hline
\end{tabular}
\end{center}
\end{table}

\subsection{Simulation results} 
To evaluate the performance of proposed mechanism in different conditions, we define the following four scenarios on the discussed simulation settings.

\subsubsection{All honest networks} 
\begin{figure}
\begin{center}
    \begin{subfigure}[t]{0.37\textwidth}
        \includegraphics[width=\textwidth]{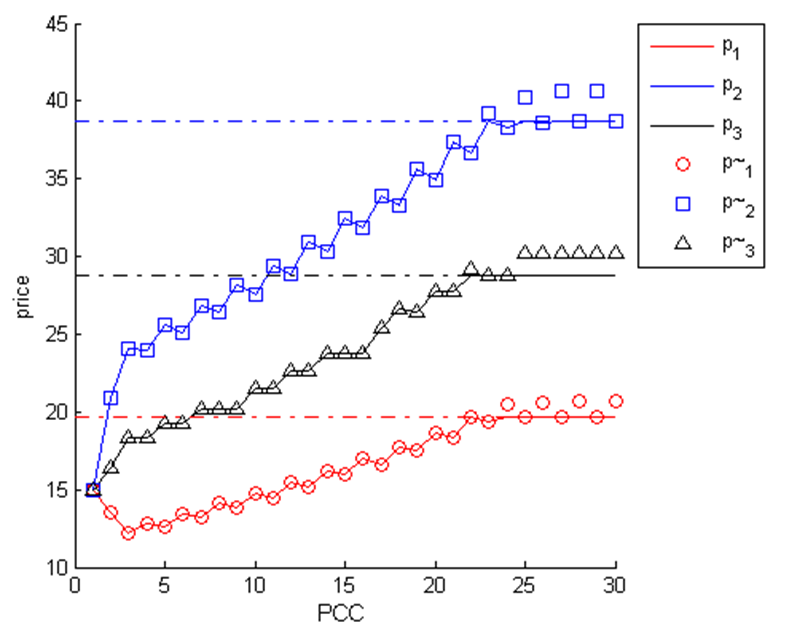}
        \caption{}
    \end{subfigure}
    \begin{subfigure}[t]{0.37\textwidth}
        \includegraphics[width=\textwidth]{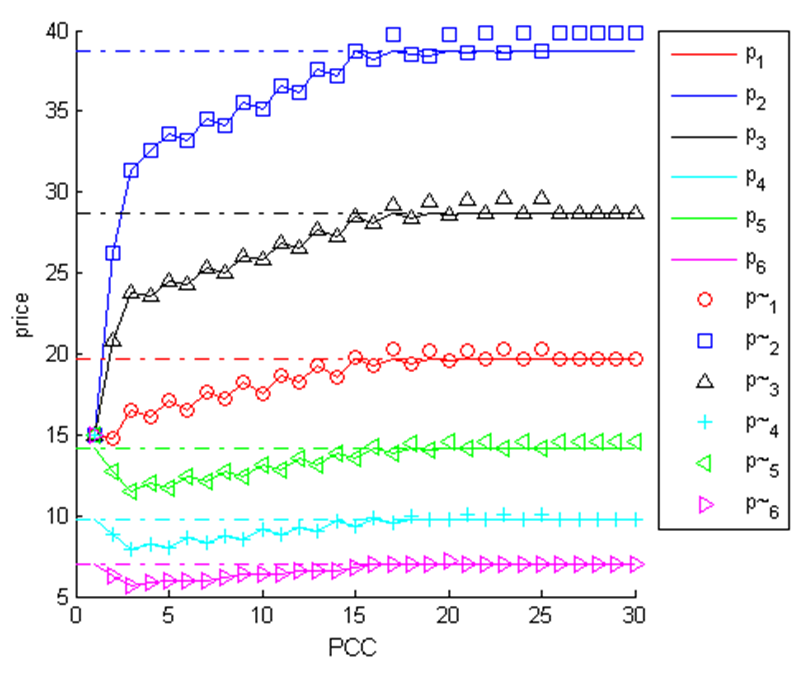}
        \caption{}
    \end{subfigure}
\caption{\footnotesize Simulation results for scenario 1, on Simulation setting 1 and 2. All WNPs are honest. The price of services converge to real marginal costs which is the private knowledge of WNPs.}\label{fig:7}
\end{center}
\end{figure}

Results of this scenario are presented in Figure \ref{fig:7}. The initial price ceilings are initiated randomly that are far from the real marginal costs and after some iterations, the regulator allows the WNPs to increase their prices until the prices meet their steady state which is near equal to real marginal costs. The parameters in our simulations are $\xi=1.05$, $\gamma=0.9$ and $\beta=2.0$. $\xi=1.05$ enables the WNPs to increase their prices. Results on both settings show that CSPC will not harm honest WNPs. In Figure \ref{fig:7} (a),(b), prices of all WNPs are converged to their real marginal cost which is the WNPs' private knowledge and the sum of absolute error (difference between real marginal cost and market price) is near zero for both settings. However, WNPs have been allowed to raise prices and since they are honest, the announced prices are at most the WNPs' marginal cost.

\subsubsection{Only WNP 1 is honest and other WNPs use the maximum allowed price} 
\begin{figure}
\begin{center}
    \begin{subfigure}[t]{0.37\textwidth}
        \includegraphics[width=\textwidth]{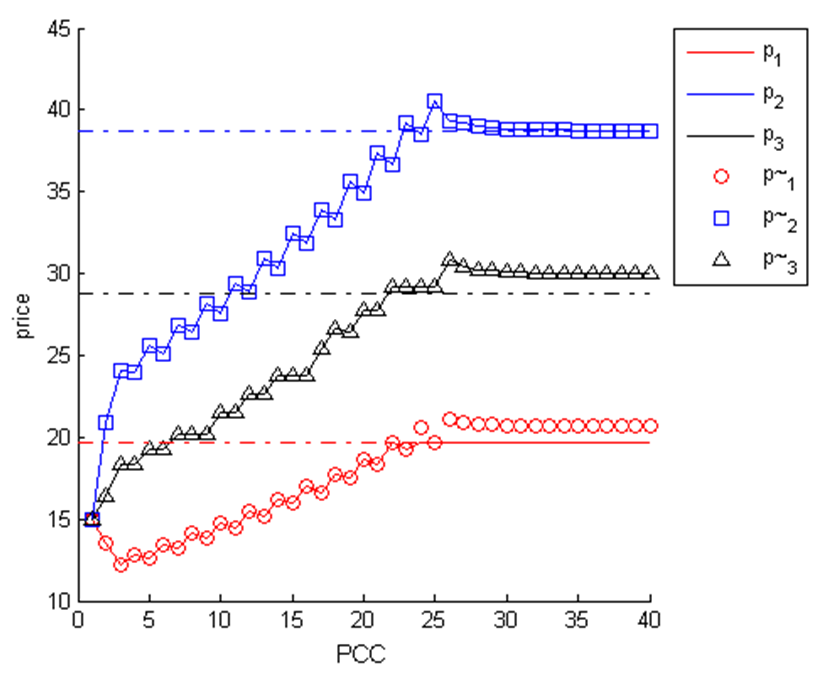}
        \caption{}
    \end{subfigure}
    \begin{subfigure}[t]{0.37\textwidth}
        \includegraphics[width=\textwidth]{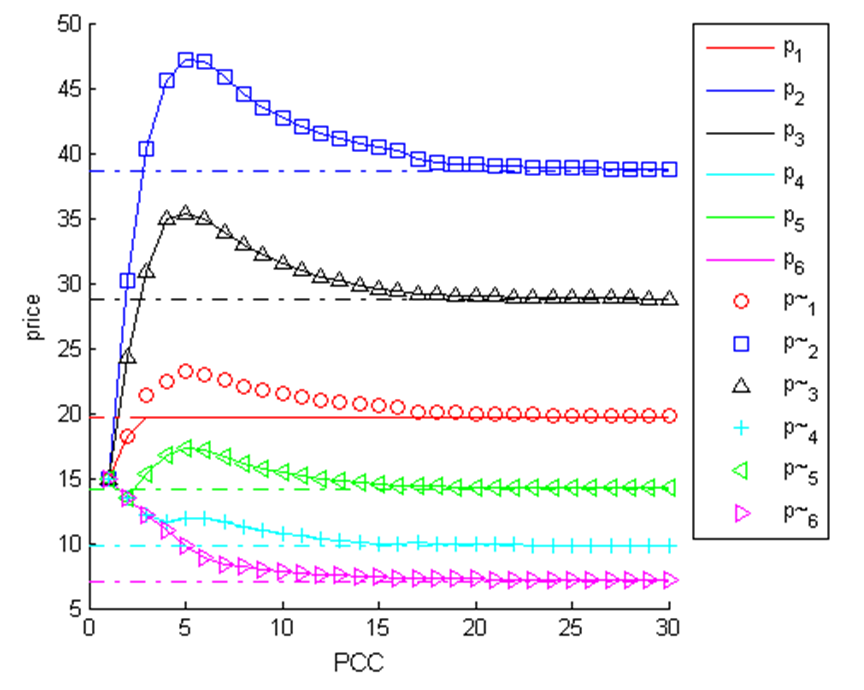}
        \caption{}
    \end{subfigure}
\caption{\footnotesize Simulation results for scenario 2, on setting 1 (a) and setting 2 in (b). Only WNP 1 is honest and other WNPs announce their price equal to maximum allowed.}\label{fig:8}
\end{center}
\end{figure}

Simulation results for this scenario are shown in Figure \ref{fig:8}. Here, we want to test that ``Can  CSPC approach control the prices if only one WNP is honest?". Hence we suppose only one of our WNPs (e.g. WNP 1), is honest and it uses its marginal cost for pricing although the regulator allows it to increase its payoff more than its marginal cost which is its private knowledge. Simulation results show that for convergence of system, it is enough to have at least one honest WNP. In Figure \ref{fig:8}(a) we study scenario 2 on setting 1. Here, after the price incremental phase until iteration 25, we find that the regulator controls the unfair WNPs 2 and 3 and reduces their price ceiling until their price converge to their marginal costs which is the private knowledge of WNPs.

Also this scenario on setting 2 shows that after the price incremental phase until iteration 4 (until WNP 1 meet its real marginal cost), the regulator is able to detect unfair WNPs and it can limit their prices which is enhanced in the following iterations meaning that after iteration 3, the only honest WNP is WNP 1 and other WNPs are correctly detected as unfair WNPs. In both settings/figures no divergence after this point is observed.
%\begin{figure}
%\begin{center}
%\includegraphics[width=.85\columnwidth]{pics/Fig009.png}
%\end{center}
%\caption{\footnotesize Absolute Sum of error for Scenario \#2 on Dataset 2.}\label{fig:9}
%\end{figure}

\subsubsection{WNP 1 and WNP 2 are honest initially; after some iterations, WNP 2 starts advertising max allowed price the same as other WNPs} 
\begin{figure}
\begin{center}
    \begin{subfigure}[t]{0.37\textwidth}
        \includegraphics[width=\textwidth]{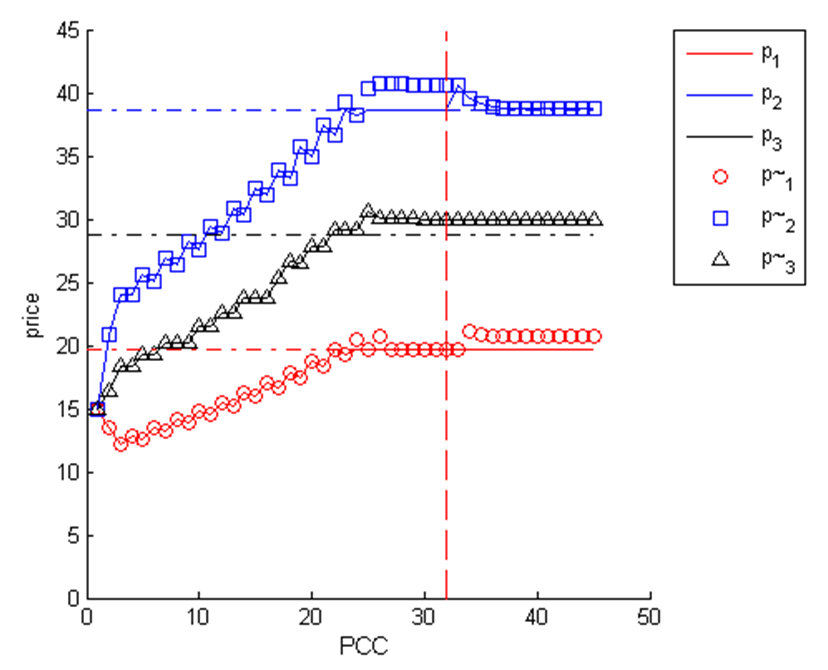}
        \caption{}
    \end{subfigure}\\
    \begin{subfigure}[t]{0.37\textwidth}
        \includegraphics[width=\textwidth]{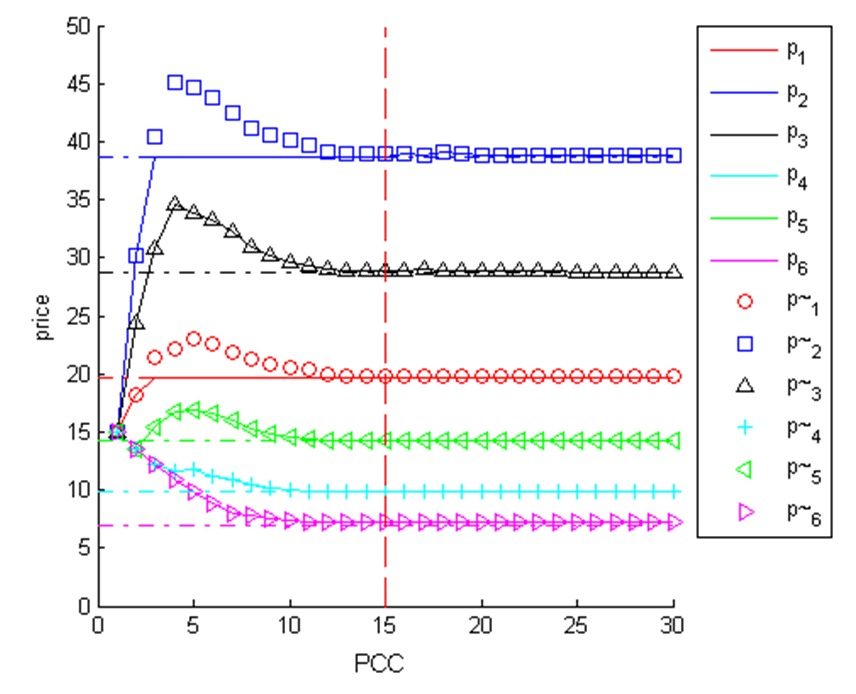}
        \caption{}
    \end{subfigure}
\caption{\footnotesize Simulation results for scenario 3, on settings 1 and 2. WNPs 1 and 2 are honest and other WNPs, announce their price as maximum allowed. a) WNP 2 changes its behavior after iteration 32 and starts to advertise unfair prices and in b) WNP 2 in iteration 15, starts to be unfair.}\label{fig:10}
\end{center}
\end{figure}

This is a test of CSPC's ability in detecting the behavioural change in the WNPs and adapting to it. In Figure \ref{fig:10}(a), WNP 2 is honest until iteration $32$, we find that the algorithm detects WNP 2 as an honest WNP and allows it to announce its price more than current price (e.g. $\tilde{p}_2=41.05$\$ while $MC_2=38.68$\$) but after iteration $32$, the behaviour of WNP 2 changes and announces the price $p_2=41.05$\$ which is its maximum allowed price. In this situation $L_2<S_2$ and using (\ref{Eq_12}), the regulator detects the network 2 as an unfair WNP; so, the regulator reduces its price ceiling $\tilde{p}_2^{new}=41.05*L_2/S_2$ . After a few iterations of price irregularities, again in iteration $35$, prices are controlled and system converges. We observe a similar situation in setting 2, Figure \ref{fig:10}(b), where the change of one WNP's behaviour from honest to unfair does not affect the system in presence of another honest WNP.

\subsubsection{Only WNP 1 with probability $\sigma$, decide to be honest} 
\begin{figure}
\begin{center}
    \begin{subfigure}[t]{0.37\textwidth}
        \includegraphics[width=\textwidth]{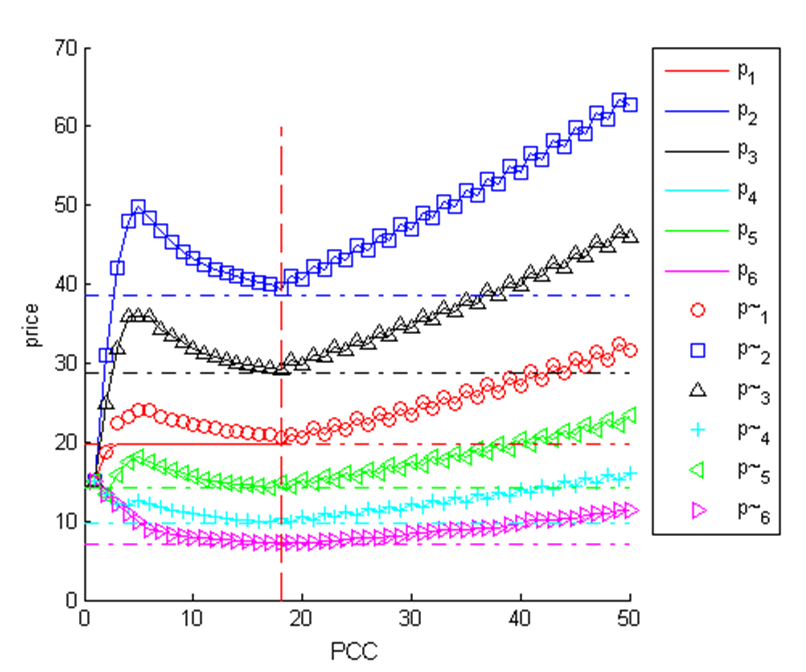}
        \caption{}
    \end{subfigure}\\
    \begin{subfigure}[t]{0.37\textwidth}
        \includegraphics[width=\textwidth]{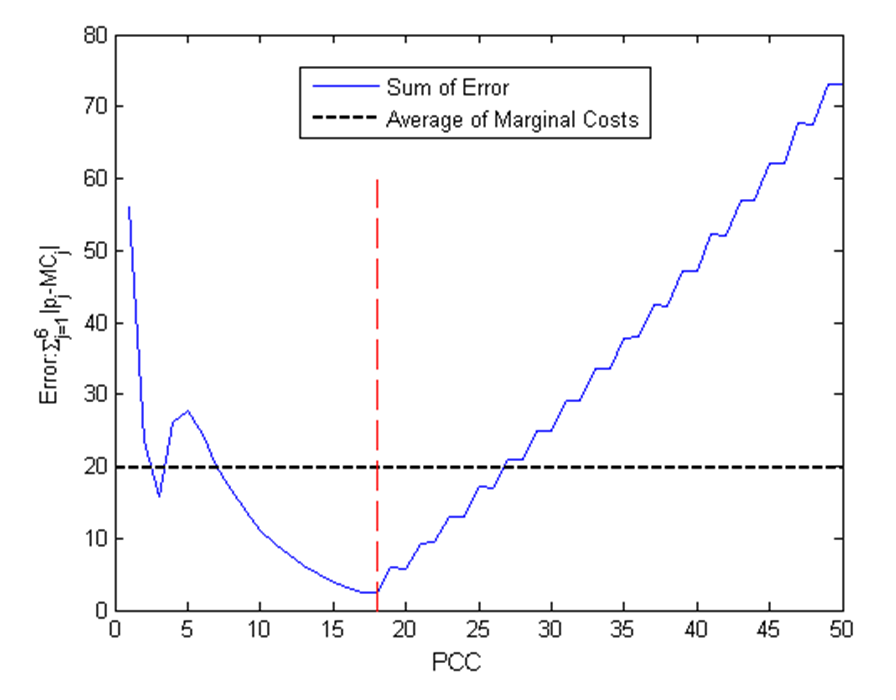}
        \caption{}
    \end{subfigure}
\caption{\footnotesize Result of simulations for scenario 4 when there is no honest WNP and all prices diverge.}\label{fig:11}
\end{center}
\end{figure}

This scenario wants to simulate a situation where all WNPs announce higher allowed price and none of the WNPs is honest. As presented in Figure \ref{fig:11}(a), until iteration 18, WNP 1 is honest and all prices are moderated by the regulator, after iteration 18, when WNP 1 changes its behavior, we find that prices are increased and there is no constraint to control the prices. In CSPC which is based on crowdsourcing, when all WNPs are unfair, the regulator does not have any tool to categorize unfair WNPs because clients send their crowdsource information according to average price which is always increasing. In Figure \ref{fig:11}(b), the sum of absolute error is near zero in iteration 18, where WNP 1 changes its behaviour. Since then, as iterations continue, the prices increase and the error of system increases continuously. 

\begin{figure}
\begin{center}
    \begin{subfigure}[t]{0.37\textwidth}
        \includegraphics[width=\textwidth]{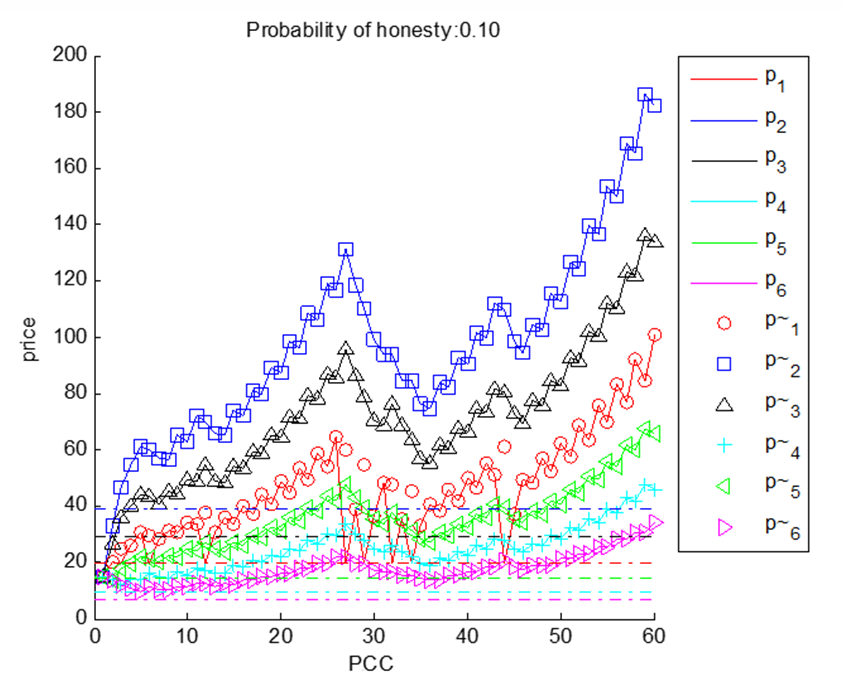}
        \caption{}
    \end{subfigure}
    \begin{subfigure}[t]{0.37\textwidth}
        \includegraphics[width=\textwidth]{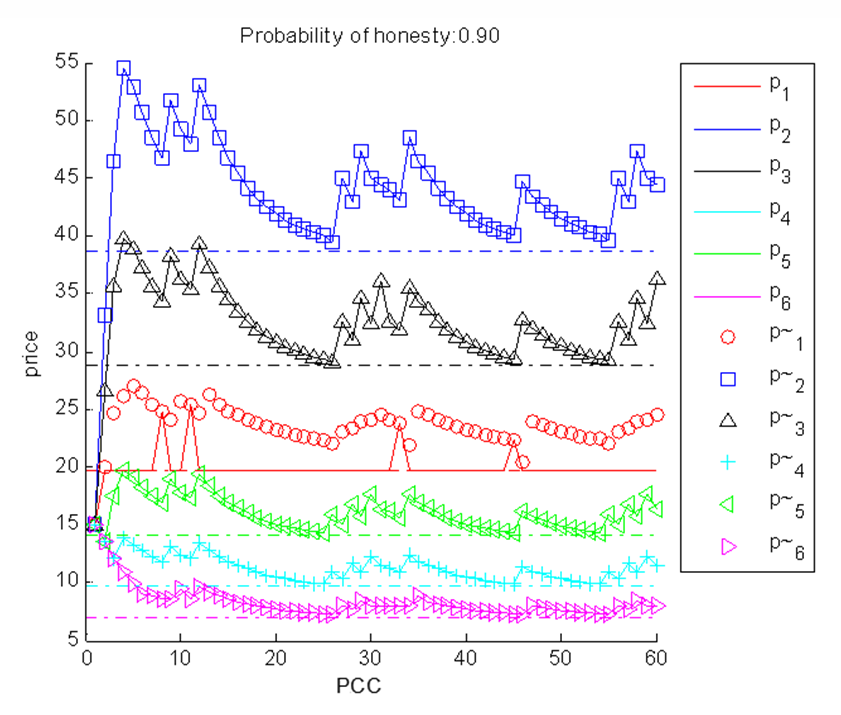}
        \caption{}
    \end{subfigure}
\caption{\footnotesize Result of simulations for scenario 4, for different probability of honesty for WNP 1 while all other WNPs are unfair. a) $\sigma=0.1$, b) $\sigma=0.9$.}\label{fig:12}
\end{center}
\end{figure}

\begin{figure}
\begin{center}
    \begin{subfigure}[t]{0.37\textwidth}
        \includegraphics[width=\textwidth]{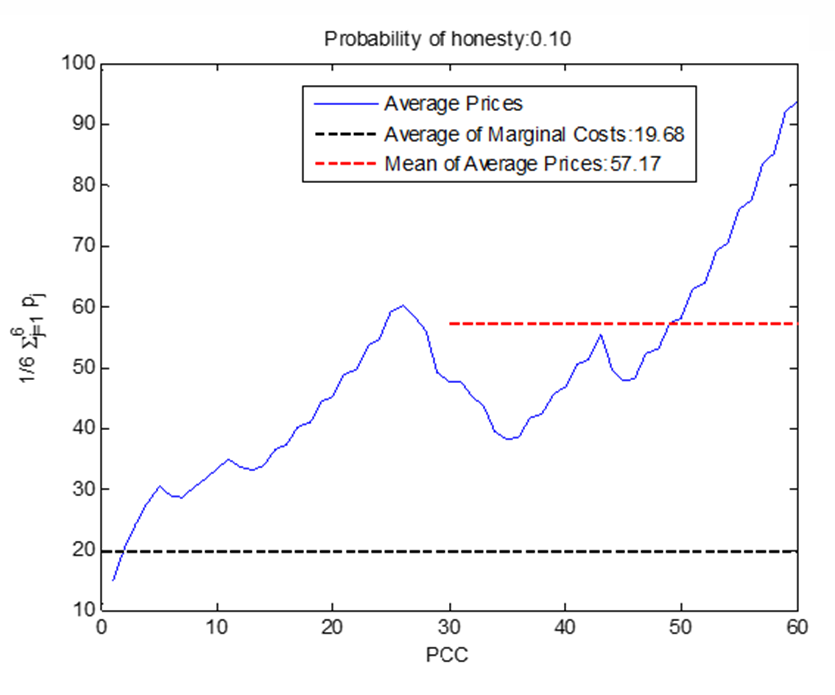}
        \caption{}
    \end{subfigure}
    \begin{subfigure}[t]{0.37\textwidth}
        \includegraphics[width=\textwidth]{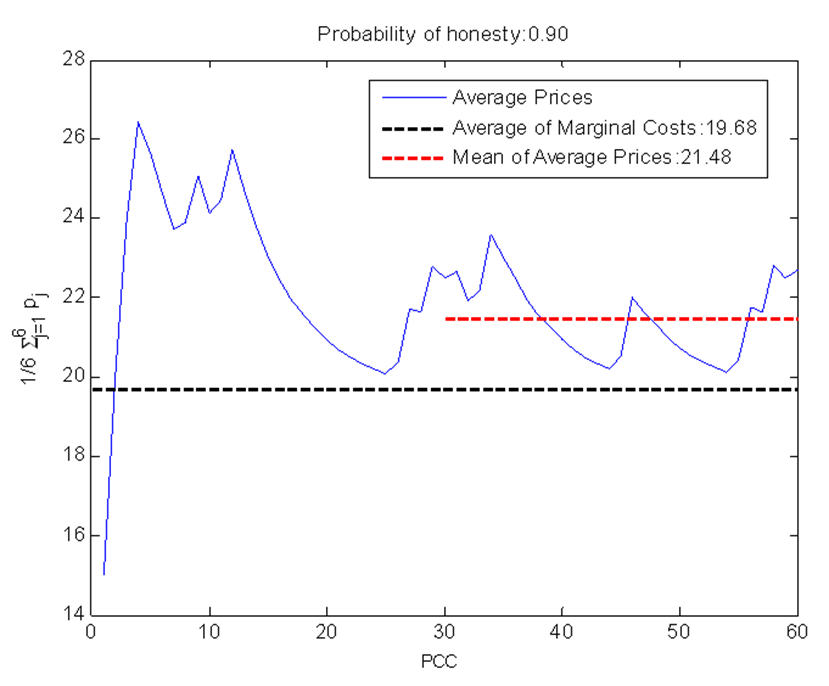}
        \caption{}
    \end{subfigure}
\caption{\footnotesize Comparison of average prices and mean of marginal costs for scenario 4, for different probability of honesty for WNP 1 while all other WNPs are unfair. a) $\sigma=0.1$, b) $\sigma=0.9$.}\label{fig:13}
\end{center}
\end{figure}

In real world, we may not find that if a WNP is always honest. In scenario 4, we want to consider a case where all WNPs are unfair and only one WNP (e.g. WNP 1) is honest with probability $\sigma$. Can CSPC still control the prices in this scenario? Figure \ref{fig:12} shows the prices for WNPs when $\sigma=0.1$ and $0.9$. In about $30$ first PCCs, system has its initial incremental of prices; so for better evaluation, we consider the mean of latest PCCs as the mean of price of specified $\sigma$ and for a better sense, Figure \ref{fig:13}, shows the average of prices $\frac{1}{6} \sum_{j=1}^6 p_j$  in each PCC for $\sigma=0.1$ and $0.9$. 

Figure \ref{fig:12}(a), shows the prices of WNPs when $\sigma=0.1$, in iterations 5,11,27,29,31,33 and 44, WNP 1 is honest but in other PCCs, it is unfair. Simulation results shows that the accumulated of surplus price is too much and few correction of prices aimed from honest pricing of WNP 1 is not enough for price controlling. Figure \ref{fig:13}(a), shows that the average of prices in the last 30 PCCs, is about $57.17$ which is about $2.91$ times more than real marginal. 
In Figure \ref{fig:12}(b), when $\sigma=0.9$, we see that only in four instances in 60 PCCs, WNP 1 is unfair and hence, the prices are well controlled and announced prices are near to the real marginal costs. The fluctuations in the Figure are due to the behaviour change of the WNP which introduces disturbance to the system. The mean of average prices in the last 30 PCCs of scenario 4 on setting 2 is equal to $21.48$ that is shown in Figure \ref{fig:13}(b) while the mean of real marginal costs is $19.68$ this means when network 1 is honest in 90\% of PCCs, the average of announced prices has about $9.1\%$ of error.

\section{Conclusions and future directions}
In this paper, we presented a three-tier game model where the regulator, WNPs and clients are the agents of our model. WNPs are licensed to use spectrum and to control the prices. The regulator, in middle time periods, adjusts the maximum price for each WNP to control the prices. The regulator strategy to adjust the maximum prices is evaluating the current load of networks and the crowdsourcing information gathered from wireless clients. WNPs which announce the price as much as  their marginal cost are named honest WNPs. The regulator can decide either a WNP is honest or not using the crowdsourced data from clients. We showed that clients can estimate the portion of prices near real portion of marginal costs. WNPs which are detected as honest are rewarded to increase their price and other WNPs are punished. In our simulations, we investigated our proposed model considering different scenarios. We showed that the best case scenario is where all WNPs are honest and the system converges easily. We also showed that the system converges if there is only one honest WNP. The worst case scenario is where there is no honest WNP and the CSPC mechanism cannot detect unfair WNPs and does not converge. We also investigated the probabilistic behaviour of WNPs. We showed that the mechanism has low error is finding fair prices if only one WNP is honest most of the time. Our findings are very promising for real world applications since it is very much possible that there exists one WNP which is honest most of the time because of its marketing policies and/or financial reasons. In our future works we will investigate detection of unfair pricing when there is no honest WNP.

%in which all WNP were honest, only one network was honest, and one network was potentially honest with a probability. In all scenarios we showed that the system converges, honest networks are not harmed, and unfair pricing is detected. Our results and analysis showed that unfair pricing is not detected only if all the networks are not honest. We will investigate this scenario in our future works. 

%\balance   
\bibliographystyle{IEEEtran}
\bibliography{MyRef}

\end{document}